\documentclass[10pt]{article}
\pdfoutput=1
\usepackage{jheppub,amsmath,amssymb}
\usepackage{enumerate}
\usepackage{bm}
\usepackage{bbm}
\usepackage[usenames,dvipsnames]{xcolor}
\usepackage{graphics}
\usepackage{tikz,todonotes}
\usepackage[utf8]{inputenc}
\usetikzlibrary{arrows,positioning,decorations.markings,decorations.pathmorphing,calc}

\definecolor{hyperref}{RGB}{026,028,087}

\newcommand{\stu}{{St\"uckelberg}}
\newcommand{\mpl}{M_{\rm Pl}}

\def\be{\begin{equation}}
\def\ee{\end{equation}}
\def\ba{\begin{eqnarray}}
\def\ea{\end{eqnarray}}

\def\nn{\nonumber}

\def\d{\mathrm{d}}

\def\ba{\begin{eqnarray}}
\def\ea{\end{eqnarray}}

\def\stu{St\"uckelberg }

\def\d{\mathrm{d}}
\def\mn{_{\mu \nu}}

\def\({\left(}
\def\){\right)}

\def\mpl{M_{\rm Pl}}

\def\ldiff{{\rm LDiff}}

\begin{document}

\title{Spin-2 and the Weak Gravity Conjecture}

\author[a,b]{Claudia de Rham}
\author[c]{Lavinia Heisenberg}
\author[a,b]{Andrew J. Tolley}
\affiliation[a]{Theoretical Physics, Blackett Laboratory, Imperial College, London, SW7 2AZ, U.K.}
\affiliation[b]{CERCA, Department of Physics, Case Western Reserve University, 10900 Euclid Ave, Cleveland, OH 44106, USA}
\affiliation[c]{Institute for Theoretical Physics,
ETH Zurich, Wolfgang-Pauli-Strasse 27, 8093, Zurich, Switzerland}

\emailAdd{c.de-rham@imperial.ac.uk}
\emailAdd{lavinia.heisenberg@phys.ethz.ch}
\emailAdd{a.tolley@imperial.ac.uk}

\abstract{Recently, it has been argued that application of the Weak Gravity Conjecture (WGC) to spin-2 fields implies a universal upper bound on the cutoff of the effective theory for a single spin-2 field. We point out here that these arguments are largely spurious,  because of the absence of states carrying spin-2 \stu $U(1)$ charge, and because of incorrect scaling assumptions. Known examples such as Kaluza-Klein theory that respect the usual WGC do so because of the existence of a genuine $U(1)$ field under which states are charged, as in the case of the \stu formulation of spin-1 theories, for which there is an unambiguously defined $U(1)$ charge. Theories of bigravity naturally satisfy a naive formulation of the WGC, $M_W< \mpl$, since the force of the massless graviton is always weaker than the massive spin-2 modes. It also follows that theories of massive gravity trivially satisfies this form of the WGC. We also point out that the identification of a massive spin-2 state in a truncated higher derivative theory, such as Einstein--Weyl--squared or its supergravity extension, bears no relationship with massive spin-2 states in the UV completion, contrary to previous statements in the literature.
We also discuss the conjecture from a  swampland perspective and show how the emergence of a universal upper bound on the cutoff relies on strong assumptions on the scale of the couplings between the spin-2 and other fields, an assumption which is known to be violated in explicit examples.

}

\maketitle


\section{Introduction}

The prevailing tool of modern physics is that of effective field theories (EFTs). Given an assumed hierarchy of scales, EFTs provide a complete description of the low energy dynamics to arbitrary order in an expansion in $E/\Lambda$ where $\Lambda$ denotes the scale of high energy physics. For instance, all gravitational quantum field theories should be understood as EFTs due to the non-renormalizable nature of their interactions where $\Lambda$ is at most $M_{\rm Planck}$ and may even be lower. Similarly, attempts to give low energy descriptions of particles with spin-2 or higher are EFTs, since their interactions are necessarily non-renormalizable. For massive spins, this is easily noted by performing a helicity decomposition from which it becomes clear that the interactions of lower helicity states contain increasing numbers of derivatives. In a weakly coupled theory, if we consider the 2--2 scattering amplitude for massive states which couple to the massive spin-2 particle, then the existence of at least one massive spin-2 particle $t$-channel exchange violates the Froissart bound, and perturbative unitarity can only be recovered by including an infinite tower of spins. This is why in all known weakly coupled UV completions, such as string theory, the spin-2 state is part of an infinite tower of excitations. However this argument does not in itself forbid a mass-gap between the spin-2 and higher states which would be sufficient to derive a low-energy EFT. A long standing question is whether there exists a theory of interacting spins $S \ge 2$ which is sufficiently gapped that it is possible to construct a low energy effective theory for a finite number of such spins. Stated differently, assuming an EFT for a particle of spin S and mass $m$, is the cutoff for that EFT parametrically larger than $m$, i.e.  $\Lambda \gg m$? It is well known that in the case of spin-2 particles coupled to gravity the cutoff is at most\footnote{Pushing $\Lambda$ to larger values requires a resonance of soft modes \cite{Gabadadze:2017jom}, or spontaneous breaking of Lorentz invariance \cite{deRham:2016plk}.} $\Lambda_3 = (m^2 \mpl)^{1/3}$ and this prevents a weakly coupled description of the limit $m \rightarrow 0$ for fixed $\mpl$. There are related conjectures for more general spins \cite{Rahman:2009zz,Bonifacio:2018aon}.\\

These specific questions are part of a larger class of questions of when can we embed a given gravitational EFT into a UV complete theory?  A prevailing view is that the EFTs can be divided into two main groups, often referred to as the Landscape and the Swampland. EFTs of the former group can successfully be embedded into a UV complete theory (e.g. string theory), whereas EFTs of the latter group represent an inhabitable space of theories incompatible with quantum gravity \cite{Ooguri:2006in}. From this point of view, it has become an indispensable task to comprehend whether a given EFT coupled to gravity belongs to the Swampland or lies in the safe Landscape. In recent years there have been a number of conjectures which define the division between the Swampland and Landscape. There is some evidence for these conjectured criteria from stringy constructions but rigorous proofs are still lacking. Among them, the conjectures on global symmetries, weak gravity, distance in moduli space and de-Sitter space have received quite some attention in the literature. Less conjectural, are the positivity bounds which can be proven given a minimal set of assumptions \cite{Adams:2006sv}.\\

The most well known folk-theorem relating to quantum gravity  is the statement that a consistent gravitationally coupled theory cannot have global symmetries \cite{Banks:1988yz,Banks:2010zn}. The motivation for this conjecture arises from the no-hair theorem for black holes in standard GR. An EFT with global symmetry coupled to gravity would have a spectrum of charged states. When these charged states are absorbed into a black hole, a stable charged remnant will form that has no apparent means to decay, violating entropy/information bounds. Hence, if an EFT possesses a global symmetry, it has to be broken, i.e. at best approximate, or gauged at some scale.\\

The weak gravity conjecture (WGC) builds on this general idea and implies in its strongest form that if an EFT with a $U(1)$ symmetry is coupled to gravity, then there should exist a fundamental cutoff for that EFT at the scale\footnote{This is the so-called magnetic version of the conjecture.}  $\Lambda = q \mpl$ \cite{ArkaniHamed:2006dz} where $q$ is the fundamental charge associated with the $U(1)$ gauge field. Assuming a weakly coupled UV completion this implies the existence of a particle whose mass $m$ is of order $q \mpl$, possibly accompanied by a tower of higher masses. From an EFT point a view there is naively no problem taking the limit $q \rightarrow 0$ for which the gauge symmetry becomes global, however this limit will run afoul of the previous folk-theorem. The universal cutoff $\Lambda = q \mpl$ explains why it is impossible to take the global limit. Additional motivation for this conjecture comes from the fact that all non-BPS black holes should be able to decay. This conjecture may be restated as the idea that gravitational force should always be the weakest \cite{ArkaniHamed:2006dz}.  Requiring that the gravitational force $F_g\sim m^2/(\mpl^2r^2)$ should be weaker than the repulsive electric force $F_g\leq F_e\sim q^2/r^2$, requires the parameters to satisfy $m/\mpl\leq q$.\\

In a recent work \cite{Klaewer:2018yxi} the interesting question of how the WGC and similar Swampland criteria constrain the EFT of a massive spin-2 field was raised. More specifically, it was conjectured that when a spin-2 particle of mass $m$, with self-interactions at the scale $M_W$ is coupled to gravity, then there is a universal upper bound on the cutoff of the low-energy EFT set by {$\Lambda_{\rm cutoff} \le \Lambda_m \approx  \frac{m \mpl}{M_W}$}. We review the precise arguments of the claim in section~\ref{subsec:helicity} but first discuss the overall logic of the arguments and the relations with the small mass limit of massive spin-2 fields.\\

 The arguments of \cite{Klaewer:2018yxi} apply the WGC to the $U(1)$ \stu symmetry that arises in the helicity decomposition of a massive spin-2 field {\it in the decoupling limit}.
 In this article, we investigate the validity of these assumptions and the conjecture.  In particular we focus our discussion on the following points:\\

\hspace{0.5cm}\begin{minipage}{14cm}
\begin{enumerate}
\item The arguments of \cite{Klaewer:2018yxi} rely on the crucial assumption that the scaling between the helicity modes and the sources has a smooth massless limit. In practice the scaling of the interactions of the helicity modes with sources typically involve negative powers of the spin-2 mass.
\item Ref. \cite{Klaewer:2018yxi} fails to assume a mass-gap between the light spin-2 states and the tower of states in the UV completion. Accounting for this leads to a very different conjecture.
\item Swampland conjectures from emergence lead to different conclusions that the conjecture.
\item The helicity-1 mode of the massive graviton is {\bf not a genuine vector field}. As a consequence,  there are {\bf no global $U(1)$ charges} associated with it and no foundations for the spin-2 WGC. This point is explained in section~\ref{sec:noWGC}.
\item {In section~\ref{Swampland}, we discuss the conjecture from a swampland perspective and show how the conjecture gets violated when diffeomorphism is preserved in the couplings with spin-2 fields, or only weakly broken.}
    \item In section~\ref{sec:examples}, we discuss examples of interacting spin-2 theories such as string theory, Kaluza-Klein and Einstein--Weyl--squared theory and show that they do not support the conjectured cutoff proposed in  \cite{Klaewer:2018yxi}.
\item Explicit UV completions of massive gravity (on AdS) such as those considered in \cite{Bachas:2011xa,Bachas:2017rch,Bachas:2018zmb}  do not respect the conjecture. In particular reference \cite{Bachas:2019rfq} proposes at alternative conjecture which is consistent with standard decoupling limit considerations.
\item As a byproduct of these considerations, we note that the identification of the massive spin-2 modes (ghosts) in certain higher derivative EFT truncations, with spin-2 states in the UV completion as argued in \cite{Ferrara:2018wlb} cannot be trusted. This means in particular that it is meaningless to apply spin-2 conjectures to ghostly massive spin-2 poles arising from a truncated expansion.
\end{enumerate}
\end{minipage}\vspace{0.5cm}

It is well-known that a massive spin-2 field in $4$ dimensions carries $5$ helicity modes corresponding to  two helicity-$\pm 2$ modes, two helicity-$\pm 1$ modes and a helicity-$0$ mode, while the massless counterpart only carries the two helicity-$\pm 2$ modes. If the helicity-$\pm 1$ and $0$ modes always entirely decoupled at the linear level from the helicity-2 modes and from external sources, one would expect a linear massive spin-2 field theory to smoothly continue to a massless spin-2 field theory in the massless limit $m\to 0$. However this is known not to be the case, indeed in 1970, van Dam and Veltman \cite{vanDam:1970vg} and independently Zakharov \cite{Zakharov:1970cc} proved that the small mass limit of a linear massive spin-2 theory does not smoothly connect with that of a massless spin-2 theory, a discontinuity now well-known under the name of vDVZ discontinuity.  Only two years later, Vainshtein understood and proved that for any massive spin-2 theory, operators that scale as negative powers of the mass enter non-linearly and including those is crucial to  the Vainshtein mechanism \cite{Vainshtein:1972sx}.
The essence of the vDVZ discontinuity, can indeed be phrased as the existence of operators that scale like inverse powers of mass. This is to be distinguished from the `observable'  vDVZ discontinuity which would suggest that physical observable in the massless limit of the massive theory differ from physical observables in the massless theory.
This `observable'  vDVZ discontinuity is an artefact of using perturbation theory in a regime where it is no longer valid as first pointed out by Vainshtein \cite{Vainshtein:1972sx}. \\

We expect a typical massive spin-2 field to have a cutoff at a scale comparable to $\Lambda_5=\(m^4 M_W\)^{1/5}$ \cite{ArkaniHamed:2002sp}, but in principle one can raise that cutoff to the scale $\Lambda_3=\(m^2 M_W\)^{1/3}$, which is indeed what is achieved in the context of nonlinear massive gravity in \cite{deRham:2010kj} or in the context of charged spin-2 fields in \cite{deRham:2014tga}. The precise way the cutoff scales with the spin-2 mass is relevant to understand precisely how the Vainshtein mechanism gets implemented, but in all cases the cutoff scales as a positive power of the mass and therefore vanishes in the massless limit, (unless Vainshtein redressing is taken into account) which is precisely once again the essence of the vDVZ discontinuity. It is precisely these interactions that undermine the assumption of a smooth massless limit which is implicit in the choice of scaling made in \cite{Klaewer:2018yxi}, as will be discussed in section~\ref{sec:noWGC}. Application of the swampland conjecture to massive spin-2 fields will also be discussed in section~\ref{Swampland} where we show that the conjecture derived in \cite{Klaewer:2018yxi} relies on strong implicit assumptions on how the spin-2 field couple to other sources. In particular we show how the conjecture gets violated in  known examples where diffeomorphism is respected in the coupling of the massive spin-2 with matter sources. In section~\ref{sec:examples} we discuss all the examples that have been provided as `{\it evidence}' for the conjecture and  show explicitly how those examples rely almost solely on the introduction of other ingredients, besides the massive spin-2 field and therefore bare little significance to massive spin-2 fields in general.
We end on more positive notes at the end of section~\ref{sec:examples} by reviewing a potential UV completion of massive gravity and bigravity on AdS and in section~\ref{sec:Positivity} by discussing powerful constraints that have successfully been imposed on the cutoff and operators of massive spin-2 fields EFTs. Requiring that the S-matrix remains analytic, Lorentz invariant, local and unitary at high energy results in positivity bounds on the scattering amplitudes of the considered low-energy EFT.

\section{WGC and massive Spin-2 fields}
\label{sec:noWGC}
\subsection{Helicity--1 mode of a massive spin-2 field}
\label{subsec:helicity}

We now review in more depth the precise arguments behind the claim of  \cite{Klaewer:2018yxi}
  and consider a (not necessarily gravitational) spin-2 field $\chi_{\mu\nu}$. The free field action on Minkowski spacetime is the Fierz-Pauli action \cite{Fierz:1939ix} and it is convenient to couple this to a spin-2 source (not necessarily the stress energy) $T_{\mu\nu}$ at a typical interaction scale {$M_W$} as a proxy for the interactions of the spin-2 field
\be
{\cal L}_{\rm FP} = \frac{1}{2} \chi^{\mu\nu} {\cal E} \chi_{\mu\nu} - \frac{1}{2}m^2(\chi_{\mu\nu}^2-\chi^2) +\frac{1}{M_W} \chi_{\mu\nu} T^{\mu\nu}\, ,
\ee
{where ${\cal E}$ is the Lichnerowitz operator}, ${\cal E}\chi_{\mu\nu} = \Box \chi_{\mu\nu} + \dots$. In this unitary gauge formulation, the degrees of freedom counting is less clear due to the presence of second class constraints. A {\it convenient} (but by no means required) way to make  the  degrees of freedom more manifest is to introduce \stu fields $A_{\mu} $ and $\pi$ via
\be
\label{eq:helicitydecomp}
\chi_{\mu\nu} =H_{\mu\nu} + \frac{1}{m} \partial_{\mu}A_{\nu}+  \frac{1}{m} \partial_{\nu}A_{\mu} + \frac{2}{m^2} \partial_{\mu}\partial_{\nu} \pi + \pi \eta_{\mu \nu}  \, ,
\ee
for which there is an associated $\ldiff \times U(1)$ local symmetry described by gauge parameters $\xi^{\mu}$ and $\chi$, {representing $U(1)$ and $\ldiff$\footnote{Linear diffeomorphisms, sometimes known as spin-2 gauge invariance.} respectively,} in the form
\ba
\label{eq:diff1}
&& H_{\mu\nu} \rightarrow H_{\mu\nu} + \partial_{\mu} \xi_{\nu}+\partial_{\nu} \xi_{\mu} + m \chi \eta_{\mu \nu}\\
&& A_{\mu} \rightarrow A_{\mu} - m \xi_{\mu}+ \partial_{\mu} \chi \\
&& \pi \rightarrow \pi -  m \chi \,.
\label{eq:diff3}
\ea
The scales $m$ are introduced so that the kinetic terms for the different fields are canonical (up to dimensionless factors of order unity). In particular, in the decoupling limit, $m \rightarrow 0$, the kinetic term cleanly separates into that for a massless helicity-2, helicity-1 and helicity-0 mode
\be
\lim_{m\rightarrow 0}{\cal L}_{\rm FP} = \frac{1}{2} H^{\mu\nu} {\cal E} H_{\mu\nu} -\frac{1}{2} F_{\mu \nu}^2 - 3 (\partial \pi)^2 + \text{sources} \, ,
 \ee
 {where $F_{\mu\nu}= \partial_{\mu}A_{\nu}- \partial_{\nu}A_{\mu}$ is the field strength of the helicity-1 field}.
 In this decoupling limit, the $U(1)$ gauge symmetry disentangles from the $\ldiff$ symmetry, and $\pi$ and $H_{\mu \nu}$ are both scalars under $U(1)$
\ba
&& H_{\mu\nu} \rightarrow H_{\mu\nu} + \partial_{\mu} \xi_{\nu}+\partial_{\nu} \xi_{\mu} \\
&& A_{\mu} \rightarrow A_{\mu} +\partial_{\mu} \chi \\
&& \pi \rightarrow \pi\,.
\ea
The source term on the other hand takes the form after integration by parts
\be
 \frac{1}{M_W}\chi\mn T^{\mu\nu}=\frac{1}{M_W}  \left[ H_{\mu\nu} T^{\mu\nu} - {\frac{2 A_{\mu} \partial_{\nu}T^{\mu\nu}}{m} }+ \pi T +2 \pi \frac{\partial_{\mu}\partial_{\nu} T^{\mu\nu}}{m^2} \right] \, .
\ee
Since for a massless spin-2 field gauge invariance requires $\partial_{\nu}T^{\mu\nu} =0$ and since the mass term will break this conservation, then it is `natural' to assume $\partial_{\nu}T^{\mu\nu} $ vanishes as some positive power of $m$. This precise behaviour will depend on the theory as we elucidate further below. The minimal choice is the one made by Schwinger \cite{Schwinger:1970xc} defined so that in the limit $m \rightarrow 0$ all source interactions are preserved:
\be
 \partial_{\nu}T^{\mu\nu}=-\frac{m}{2}  J^{\mu} \, , \quad \partial_{\mu} J^{\mu} =  -m (K-T) \, , \quad \text{(Schwinger)}
\ee
so that the source interactions take the finite form
\be
 \frac{1}{M_W}  \left[ H_{\mu\nu} T^{\mu\nu} + A_{\mu} J^{\mu} +  \pi  K \right] \, .
\ee
In practice as we discuss below, this will require some specific choice of scaling of $M_W$ or of the interactions that make up the real sources. For exactly conserved sources, we have $K=T$ and we recover the van Dam-Veltman-Zakharov (vDVZ) discontinuity\footnote{It is clear from Schwinger's analysis that the essence of the discontinuity was already understood in \cite{Schwinger:1970xc}. } \cite{vanDam:1970vg,Zakharov:1970cc}.\\

By contrast the \textbf{\emph{choice}}  made in \cite{Klaewer:2018yxi} is
\be
\partial_{\nu}T^{\mu\nu} = -\frac{m^2}{2} \tilde J^{\mu} \, , \label{stessenergyassumption}
\ee
for which the source term is
\be
 \frac{1}{M_W}\chi\mn T^{\mu\nu}= \frac{1}{M_W}  H_{\mu\nu} T^{\mu\nu} +  \frac{m}{M_W}A_{\mu} \tilde J^{\mu}+  \frac{1}{M_W}  \pi T -  \frac{1}{M_W} \pi \partial_{\mu} \tilde J^{\mu} \, .
\ee
We further expect a suppression of $ \partial_{\mu} \tilde J^{\mu} $ at least of the form $\partial_{\mu} \tilde J^{\mu} = -m \tilde K$ for which in the limit $m\rightarrow 0$ (for fixed $M_W$) the source for the helicity-1 mode vanishes
\be
 \frac{1}{M_W}\chi\mn T^{\mu\nu} \xrightarrow{m\to 0} \frac{1}{M_W}  H_{\mu\nu} T^{\mu\nu} + \frac{1}{M_W}  \pi T   \, .
\ee
This is the well known result that the force exchanged between two conserved sources by a massive spin-2 mode is dominated by the helicity-2 and helicity-0 modes.\\

The key argument of \cite{Klaewer:2018yxi} is that the smallness of the interaction for the helicity-1 mode $ \frac{m}{M_W}A_{\mu} J^{\mu}$ may be interpreted as effectively saying that the associated gauge coupling is suppressed as {$g_m \sim m/M_W$}. Assuming the validity of this, if we now couple this theory to a massless graviton, whose own interaction scale is $\mpl$, then the WGC would seem to claim that the effective cutoff of the theory is at most
\be
\label{eq:conjecture}
 \Lambda_m \sim g_m \mpl = \frac{m \mpl}{M_W} \, .
\ee
There are three essential problems with this argument:
\begin{itemize}
\item There are no states, neither fundamental nor composite particles, charged under this `fake' $U(1)$ symmetry. Since there is no global limit of the $U(1)$ symmetry, there is no known tension in coupling this theory to gravity as we shall explain in details in sections~\ref{subsec:noU1} and \ref{subsec:noCharges}.
\item In general the $m$ scaling of $\partial_{\nu}T^{\mu\nu} $ is such that interactions blow up as $m \rightarrow 0$ for fixed $M_W$, meaning the associated current become large in this limit as we explain in sections~\ref{subsec:scaling} and \ref{subsec:scaling2}. The identification of a gauge coupling $g_m \sim m/M_W$ is largely meaningless.
\item The argument itself assumes that the mass scale from the interactions is also $m$ as would be the case of the massive spin two state sits in an infinite tower of states spaced by the same scale. This assumes in advance that the cutoff is of order $m$ due to the existence of these additional states at the scale of order $m$.
\end{itemize}

\subsection{Assuming a Mass Gap $m \ll \Lambda_t$.}

We will expand on the last point first. In order to consider a low energy effective theory with at least one massive spin-2 state and a finite number of other states, in a typical UV completion which may contain an infinite number of states, it is crucial that the mass of the spin-2 state lies well below the lowest scale in the infinite tower that defines the UV completion, which we denote as $\Lambda_t$ (for `tower'). If this is not the case, then the low energy effective theory is not under control. Standard Kaluza-Klein theories do not provide an example of a UV completion of massive spin-2 effective theories precisely because there is in general no significant gap, for Kaluza-Klein we quite generically have $\Lambda_t \sim m$. If we discard WGC considerations and focus entirely on swampland considerations, bounds on the coupling will come from the infinite tower of states. It is then more natural to suppose that it is these interactions will be the ones of concern, i.e. so that it is more natural to identify
\be
\partial_{\nu}T^{\mu\nu} = -\frac{\Lambda_t^2}{2} \tilde J^{\mu} \, .
\ee
Indeed this is consistent with the argumentation we present in section \ref{Swampland}.
In a standard weakly coupled UV completion, it will be this infinite tower whose role is to cure the problems of the low energy effective theory and so we identify $\Lambda_m \sim \Lambda_t$. We then obtain $g_m \sim \Lambda_t^3 \mpl/M_W$ and so we identify
\be
\Lambda_m =\Lambda_t \sim \frac{\Lambda_t^2 \mpl}{m M_W} \, .
\ee
Hence we infer the cutoff of the low energy EFT to be
\be
\Lambda_t \sim \frac{m M_W}{\mpl} \, .
\ee
Since by assumption $m \ll \Lambda_t$, then we infer $M_W \gg \mpl$, which is entirely distinct from the previous argument. Both arguments are essentially too simplistic as we discuss in section \ref{Swampland}, since they neglect to account for the fact that in an explicit UV completion of a theory with a light spin-2 state, interactions of the spin-2 with states in the tower will necessarily be suppressed by $m$ due to the diffeomorphism/spin-2 gauge invariance that arises in the limit $m \rightarrow 0$.

\subsection{No Global $U(1)$ charges}
\label{subsec:noU1}

\paragraph{The case of a `Fake' $U(1)_A$:}  With regards to WGC considerations, the first point is that $A_{\mu}$ is not a true $U(1)$ gauge field.
If it were, it would be possible to consider states, (either fundamental states or composite ones), charged under this $U(1)$ symmetry. If we could construct a black hole carrying such a global charge, then we would be led to the usual conundrum that leads to the WGC {\cite{ArkaniHamed:2006dz}}. However  in the case considered here, the field $A_\mu$ introduced in \eqref{eq:helicitydecomp} is {\bf not} a gauge field that has been introduced to gauge a global symmetry down to a local one.

To see why there are no states charged under this `fake' $U(1)$ (what we shall call the $U(1)_A$), consider first the decoupling limit theory $m \rightarrow 0$ for which $A_{\mu}$ transforms as a $U(1)$ gauge field. In {\it that decoupling limit} we may for example {introduce some complex} spin-0 source $\Phi$ charged under this gauge field, i.e. described by a Lagrangian
\be
{\cal L}_{\phi}=|\partial_{\mu} \Phi -i q A_{\mu} \Phi|^2-V(|\Phi|^2) \, .
\ee
As we move away from the decoupling limit, it is natural to ask whether there is a Lagrangian for which $\Phi$ {would couple to all of the spin-2 field components $H_{\mu\nu}$, $A_\mu$ and $\pi$, which would reduce} in the decoupling limit to this form. It is however easy to see that there is no such local Lagrangian, precisely because of the modified form of the gauge transformations $A_{\mu} \rightarrow A_{\mu} + m \xi_{\mu}+ \partial_{\mu} \chi $ for $m \neq 0$. For instance, varying ${\cal L}_{\phi}$ we find
\be
\delta {\cal L}_{\phi} = - i q m \xi^{\mu} ( \Phi^* \partial_{\mu} \Phi-  \partial_{\mu} \Phi^*  \Phi) + \dots
\ee
In order to construct a gauge invariant Lagrangian, we {would need} to add at $O(m)$ a cubic interaction whose transformation cancels this $O(m)$ term. But such an interaction must necessarily include terms which transform with $\xi^{\mu}$ without derivatives. There are no such combination (other than $A_{\mu}$ itself) unless we force ourselves to consider non-local ones, which is equivalent to demanding the presence of additional fields. Indeed forcing ourselves into non-local combinations, the only combination other than  $A_{\mu}$ itself that transforms as $\xi^{\mu}$ without derivatives is
\be
\frac{1}{\Box}\partial_{\mu}(H^{\mu\nu}- \frac{1}{2} \eta^{\mu\nu}H)- \frac{1}{\Box} \partial^{\nu} \pi  \rightarrow \frac{1}{\Box}\partial_{\mu}(H^{\mu\nu}- \frac{1}{2} \eta^{\mu\nu}H) - \frac{1}{\Box} \partial^{\nu} \pi + \xi^{\nu}
\ee
and we would necessarily be forced into a non-local interaction. The appearance of non-locality should be interpreted as arising from having integrated out some massless degrees of freedom, but if such a degree of freedom were introduced, it would necessarily give rise to a light ghost. Specifically, we could imagine introducing a vector field $B^{\mu}$ which transforms as $B^{\mu} \rightarrow B^{\mu}-\xi^{\mu}$. We can construct a local Lagrangian by means of a Lagrange multiplier $\lambda^{\mu}$ by adding a term
\be
\lambda^{\mu}\left( \Box B_{\mu} -   \partial_{\mu}(H^{\mu\nu}- \frac{1}{2} \eta^{\mu\nu}H)+  \partial^{\mu} \pi \right)\,,
\ee
and then couple the charged scalar to
\be
{\cal L}_{\phi}=|\partial_{\mu} \Phi -i q ( A_{\mu}- m B_{\mu}) \Phi|^2-V(|\Phi|^2) \, .
\ee
This is gauge invariant under the full $\ldiff \times U(1)$ symmetry but the problem is that the Hamiltonian is unbounded from below in its kinetic structure signalling a light ghost. We may also choose to add a separate kinetic term for the true $U(1)$ gauge field $A_{\mu}- m B_{\mu}$ but {in doing so} we are really introducing a new independent spin-1 field $B'_{\mu} = A_{\mu}- m B_{\mu}$ whose $U(1)_B$ symmetry is independent of the original \stu one. In other words, the only way to apply the WGC on massive spin-2 fields in the sense of \cite{Klaewer:2018yxi} would be to force an extra  sector on the massive spin-2 field so as to oblige it to carry a genuine $U(1)_B$ symmetry and being able to constrain that $U(1)_B$ sector.
Stated yet differently, the WGC does not actually constrain the massive spin-2 EFT but rather an external $U(1)_B$ sector one may choose (or not) to couple to the massive spin-2 EFT. Of course out of any massive spin-2 EFT one can always add on an external $U(1)_B$ sector (and sometimes this extra sector enters naturally) as is the case in the examples provided in section~\ref{sec:examples}, but those are genuine implementation of the WGC onto the $U(1)_B$ sector and have very little to do with the existence of a massive spin-2 field and have certainly nothing to do with the implementation of the WGC on the massive spin-2 EFT. \\

This result is entirely clear in unitary gauge where there is {\bf no $U(1)$ symmetry, and hence no global remnant of this $U(1)$ symmetry}. In this sense the current $J^{\mu}$ cannot be meaningfully associated with any charged state which is conserved in a global limit, as would be necessary in order to run the WGC arguments. This is not to say that $J^{\mu}$ is zero, but rather that it arises only from interactions that specifically {come} from $m \neq 0$ and there is no meaningful conservation law associated with $J^{\mu}$ in any massless limit. This situation is very different to the spin-1 \stu theories {\cite{Reece:2018zvv}. In this case,} there exists a genuine $U(1)$ gauge field to which charged particles may couple. Furthermore there exists a global limit of such a theory.\\

\paragraph{The case of a genuine $U(1)_B$:} The arguments presented above are different from what happens in a theory with  a {\it genuine} gauge symmetry, even if it is broken by a mass term.
For example, give a \stu  theory for spin-1 with some interactions
\be
{\cal L}_B = - \frac{1}{4} F_{\mu\nu}^2 - \frac{1}{2} m^2 B_{\mu}^2 + \lambda (B_{\mu}^2)^2\,.
\ee
The \stu  procedure amounts to defining $B_{\mu} = \tilde B_{\mu} + \frac{1}{m} \partial_{\mu} \pi$ where again the $1/m$ is introduced to set normalization. We may choose to couple this to a charged scalar in the same form ${\cal L}_{\phi}=|\partial_{\mu} \Phi -i q  \tilde B_{\mu} \Phi|^2-V(|\Phi|^2) $. We may then consider a decoupling limit $m \rightarrow 0$, $q \rightarrow 0$, keeping $\Lambda/m^4= 1/\Lambda^4$ fixed for which the Lagrangian becomes
\be
{\cal L}= - \frac{1}{4} F_{\mu\nu}^2 - \frac{1}{2} (\partial \pi)^2 + \frac{1}{\Lambda^4} (\partial \pi)^4+|\partial_{\mu} \Phi|^2-V(|\Phi|^2)\,.
\ee
This gives us a theory with a genuine global $U(1)_B$ symmetry, for which $J^{\mu}$ is conserved, despite the fact that the local $U(1)_B$ symmetry was spontaneously broken. The naive cutoff is $\Lambda$. Coupling this to gravity would be in tension with the WGC, thus, placing a lower limit on $q$, i.e. requiring the true cutoff of the EFT to be\footnote{A more refined and restricted set of conjectures are given for \stu fields in \cite{Reece:2018zvv}.} $\Lambda_{\rm cutoff} = {\rm Min}(\Lambda, q \mpl)$. We emphasize however that this situation is very different from that considered in the case of the `fake' $U(1)_A$ presented in section~\ref{subsec:helicity} in the case of massive spin-2 fields, where there is no separate genuine $U(1)_B$,  hence no notion of global $U(1)$ charges.

\subsection{Absence of charges for static sources}
\label{subsec:noCharges}

The fact that no matter fields transform under this `fake' $U(1)_A$ symmetry has another importance consequence: for a static source we necessarily have $Q=0$ nonlinearly.
Let us define a dimensionless charge $Q$ by
\be
Q = \frac{1}{M_W}\int \d^3 x J^0 = -\frac{1}{M_W}\int \d^3 x \partial_{\mu} T^{\mu 0}= -\frac{\d}{\d t} \int \d^3 x T^{00}\, .
\ee
Consider a static source, e.g. that of a point particle for which the only non-zero component of the stress energy is $T^{00} = M_s \delta^3(x)$, at zeroth order in $m$. At leading order $Q=0$ since the mass is conserved, however we might expect this to be non-zero at higher orders in $m$ due to the non-conservation of the stress energy. This violation of conservation is theory specific but will take the schematic form
\be
\partial_{\mu} T^{\mu 0} = F^{0}[\partial_a,  \chi_{\mu\nu},T_{\alpha\beta}] \, .
\ee
From index contractions, there must be an odd number of $0$'s on the right hand side. Given any equation for a tensor $\chi_{\mu\nu}$ alone, with a source for which $T^{0i}=0$, and given the assumed static nature $\partial^0=0$ then at any order in perturbations $\chi^{0i}=0$, from which we conclude that $\partial_{\mu} T^{\mu 0} =0$ at all orders. In other words in order to make a current $J^{\mu}$ out of tensors, we need a vector, but the only vector in play is $\partial^{\mu}$ which has zero timelike component for static sources.

In order to construct a static charge, we need a {\bf genuine} $U(1)$ symmetry with the phase chosen to be dependent linearly in time. For instance a complex $U(1)$ scalar field $\Phi(x) = e^{-i \alpha t} \varphi(\vec x)$ will give a static charge density and stress energy since the time dependent phase drops out by virtue of the genuine $U(1)$ symmetry. In this way we have $\partial^0 \neq 0$ without violating the static assumption. Elementary particles acquire charges exactly in this manner through the quantization of fields in the form $\Phi(x) = e^{-i \omega t} e^{ik.x} a_k+ \dots$.

Hence, in the case of the \stu helicity-1 field with the `fake' $U(1)_A$ symmetry, there are no charged states in the presence of a static source, which is at the heart of the WGC conjecture.

\subsection{Scaling the Interactions}
\label{subsec:scaling}

Let us now further clarify the meaning of the assumption \eqref{stessenergyassumption}. To do this, it is helpful to work with a simple example. Although the spin-2 field does not have to couple to the stress energy, let us imagine that it does so, and couples to that of a massless  scalar field $\varphi$ so that the Lagrangian contains
\be
{\cal L} \supset - \frac{1}{2} (\partial \varphi)^2 + \frac{\chi_{\mu\nu}}{M_W} T^{\mu\nu}
\ee
where $T^{\mu\nu} = \partial^{\mu} \varphi\,  \partial^{\nu} \varphi - \frac{1}{2} \eta^{\mu\nu}(\partial \varphi)^2$. A naive dimensional analysis implies that at finite order the cutoff of this EFT is at most $(m^2 M_W)^{1/3}$ from the interactions of the form
\be
{\cal L} \supset \frac{\partial_{\mu} \partial_{\nu} \pi}{m^2 M_W} (\partial^{\mu} \varphi\,  \partial^{\nu} \varphi - \frac{1}{2} \eta^{\mu\nu}(\partial \varphi)^2) \, .
\ee
These cubic interactions can of course be removed with a field redefinition but will survive at higher order unless specifically chosen to cancel. Because of the interaction, the stress energy is no longer conserved and we find
\be
\partial_{\alpha} T^{\alpha \beta} = \partial^{\beta} \varphi \Box \varphi = \frac{2}{M_W} \chi_{\mu\nu} \partial^{\beta}\varphi \left[\partial^{\mu}\partial^{\nu}\varphi- \frac{1}{2} \eta^{\mu\nu} \Box \varphi \right] + \frac{2}{M_W}\partial^{\beta} \varphi \partial^{\mu} \varphi \left[ \partial^{\nu} \chi_{\mu\nu}- \frac{1}{2} \partial_{\mu} \chi\right] \, .
\ee
This is analogous to the covariant conservation equation in GR $\nabla_{\mu} T^{\mu\nu} \rightarrow \partial_{\mu} T^{\mu\nu} = -\Gamma^{\mu}_{\mu \alpha} T^{\alpha \nu}- \Gamma^{\nu}_{\mu \alpha} T^{\mu \alpha}$.

Focusing on the leading helicity-zero mode contribution on the right hand side we have schematically \cite{deRham:2011qq}
\be
\partial_{\alpha} T^{\alpha \beta}  \sim \frac{1}{m^2 M_W} \partial \partial \partial \pi T + \dots \, .
\ee
It is apparent from this form that for fixed $M_W$ this becomes arbitrarily large as $m \rightarrow 0$. This is non other than the usual vDVZ discontinuity \cite{vanDam:1970vg,Zakharov:1970cc}. The sense in which this term is $m^2$ suppressed \cite{deRham:2011qq} is for fixed $\Lambda_5 = (m^4 M_W)^{1/5}$
\be
\partial_{\alpha} T^{\alpha \beta}  \sim \frac{m^2}{\Lambda_5^5} \partial \partial \partial \pi T + \dots \, .
\ee
Since $\Lambda_5 = (m^2 M_W)^{1/5}$ is the typical cutoff for the interactions of a massive spin-2 particle without any particular structure, then what is true is if we take a decoupling limit $M_W \rightarrow \infty$, $m \rightarrow 0$ keeping $\Lambda_5$ fixed, $\partial_{\alpha} T^{\alpha \beta}$ would be suppressed by a single power of $m^2$.

Nevertheless it remains the case that there is no meaningful sense in which we can identify a gauge coupling as $g_m \sim m/M_W$. The structure of the actual interaction is entirely different \cite{deRham:2011qq}.

\subsection{Gravity as the weakest force}
\label{subsec:scaling2}

A more mundane definition of the WGC is that the force induced by anything other than the massless graviton, is stronger than that of the massless graviton, i.e. gravity is the weakest force.
Clearly this statement can only apply at distances less than the Compton wavelength of the massive states, since in the extreme IR the massless graviton is always strongest.  Let us consider two well separated nonlinear configurations A and B for $\chi_{\mu\nu}$ sourced by for example a delta function stress energy $T^{00}=  M_A \delta^3(r-r_A)+ M_B \delta^3(r-r_B)$. In general associated with the self interactions scale $M_W$ is a 'Vainshtein' radius $r_V$ of the form $r_V^{A,B}= \Lambda_n^{-1}\left( M_s/M_W\right)^{1/n} $, where $\Lambda_n^n = M_W m^{n-1}$,  which sets the scale at which the interactions of the helicity-zero mode of the spin-2 state become significant. The precise values of $n$ depend on the {concrete} interactions, see Ref.~\cite{Babichev:2013usa} for a review on the Vainshtein mechanism.

Two well known cases are the $\Lambda_5$ theories and $\Lambda_3$ theories (see \cite{deRham:2018qqo} for a distinction between these different types of theories). A $\Lambda_5$ theory of spin-2 particle is one for which the unitary gauge theory includes self-interactions of the form
\be
{\cal L}= \frac{1}{2} \chi^{\mu\nu} {\cal E} \chi_{\mu\nu} - \frac{1}{2}m^2(\chi_{\mu\nu}^2-\chi^2) + \frac{m^2}{M_W} (a_1 {\rm Tr}({\chi^{\mu}_{\nu}}^3) + a_2 {\rm Tr}({\chi^{\mu}_{\nu}}^2){\rm Tr}(\chi^{\mu}_{\nu})  + a_3 ({\rm Tr}(\chi^{\mu}_{\nu}) )^3)+\dots\, ,
\ee
where $a_i$ are order unity dimensionless coefficients. With no special choice of values for $a_i $, introducing \stu  fields this interactions lead to irrelevant interactions at the scale $\Lambda_5$
\be
\frac{m^2}{M_W} {\rm Tr}((\chi^{\mu}_{\nu})^3) \sim \frac{m^2}{M_W} (\partial \partial \pi/m^2)^3 \sim \frac{1}{\Lambda_5^5}(\partial \partial \pi)^3 \, .
\ee
The associated Vainshtein radius associated with a source of mass $M_s$ is determined by asking when the weak field approximation $\pi \sim M_s/(M_W r)$ breaks down. This is when
\be
(\partial \pi)^2 \sim \frac{1}{\Lambda_5^5}(\partial \partial \pi)^3
\ee
for which
\be
r_V  = \frac{1}{\Lambda_5} \left( \frac{M_s}{M_W} \right)^{1/5}\,.
\ee
The $\Lambda_3$ theory on the other hand is obtained by tuning the coefficients $a_i$ into the structure
\be
{\cal L} = \frac{1}{2} \chi^{\mu\nu} {\cal E} \chi_{\mu\nu} - \frac{1}{2}m^2(\chi_{\mu\nu}^2-\chi^2) + \frac{m^2}{M_W} \tilde a \epsilon^{abcd} \epsilon^{ABCD} \chi_{aA} \chi_{bB} \chi_{cC} \eta_{dD}+\dots\, ,
\ee
Again introducing \stu fields, the leading interactions are
\be
2 \tilde a \frac{m^2}{M_W} \frac{1}{m^4}  \epsilon^{abcd} \epsilon^{ABCD} H_{aA} \pi_{bB} \pi_{cC} \eta_{dD}= 2 \tilde a \frac{1}{\Lambda_3^3}  \epsilon^{abcd} \epsilon^{ABCD} H_{aA} \pi_{bB} \pi_{cC} \eta_{dD} \, .
\ee
In this case the nonlinearities become important when $\partial \partial \pi \sim \Lambda_3^3$, i.e. at the Vainshtein radius
\be
r_V  = \frac{1}{\Lambda_3} \left( \frac{M_s}{M_W} \right)^{1/3} \, .
\ee

If we consider two such configurations which are separated by a distance much larger than $r_V$ then it is meaningful to talk about the force between them. When they are closer than $r_V$ the nonlinearities extend over the region of both particles and it is difficult to infer the force between them without solving a nonlinear problem.

As we have already discussed, given these two well separated configurations, we can associate with them some charge $Q_A$ and $Q_B$ {induced} by the integral of $J^{\mu}$ over their separate nonlinear profiles via
\be
Q = \frac{1}{M_W}\int \d^3 x J^0 \, .
\ee
If the sources are truly static then $Q$ vanishes and there is no force to talk about. We can however imagine the force between two objects, e.g. molecules, galaxies, for which the internal components are moving so for which $\partial^0 \sim \epsilon \partial^i$. Given then that there must be at least one power of $\epsilon$ on the right hand side, we can perform simple dimensional analysis to get a feel for the magnitude of the charges. From our above estimate
\be
Q \sim \frac{1}{M_W}\int \d^3 x \frac{1}{m^3 M_W^2} \partial \partial \partial \pi T \, .
\ee
Since the interaction may only be treated perturbatively at distances $r \gg r_V$, a simple estimate of the charge is obtained by assuming the mass of the sources is evenly distributed over the Vainshtein radius. Thus we have
\be
Q \sim r_V^3 \epsilon \frac{1}{m^3 M_W^2} \frac{M_s}{M_W r_V^4} \frac{M_s}{r_V^3} \sim \epsilon \frac{M_s^2}{m^3 M_W^3 r_V^4} \, .
\ee
For the case of a $\Lambda_3$ theory this corresponds to
\be
Q \sim \epsilon  \left( \frac{M_s^{2}}{m M_W} \right)^{1/3} \, ,
\ee
whereas for the $\Lambda_5$ theory it is
\be
Q \sim \epsilon \left( \frac{M_s^{6} m}{ M_W^7} \right)^{1/5} \, .
\ee
Demanding the WGC hold in the sense that the force exchanged by the helicity-1 mode is larger than the force exchanged by a massless graviton, requires
\be
Q \gtrsim M_s/\mpl \, .
\ee
In the case of a $\Lambda_3$ theory then this indeed gives
\be
M_s \lesssim \frac{\epsilon^3 \mpl^3}{m M_W}\,,
\ee
and in the case of a $\Lambda_5$ theory
\be
M_s \gtrsim \frac{M_W^7}{m \epsilon^5 \mpl^5} \, .
\ee
Neither of these estimates bears any resemblance to the conjecture of \eqref{eq:conjecture} and both are clearly very sensitive to the nonlinear nature of the interactions. The WGC applied to the force exchanged by the helicity-1 states does not appear to give anything definitive, and neither should it be since there is no physical content in separating out the contribution from that mode.

In this context it may  indeed appear that the repulsive force from the helicity-1 exchange will generically be weaker than that for massless graviton exchange and indeed vanishes for static sources since there are no static charges. However this small repulsive force cannot in real terms be isolated from the much larger {attractive} force mediated by the helicity-2 and helicity-0 exchange. All that is physically relevant is the total force induced by the exchange of the spin-2 field. Note in particular that the helicity-zero attractive force $\frac{1}{M_W} \pi T $ between two $J^{\mu}$ sources will automatically dominate over the helicity-1 repulsion. As long as {$\mpl>M_W$}, which is the typical assumption, gravity (meaning the force exchanged by the massless spin-2) will remain the weakest force.

\section{Swampland Conjectures for Spin-2 from emergence}

\label{Swampland}
Related but distinct from the WGC are various `swampland conjectures' which have been applied to amongst other things scalar field theories which have no particular gauge symmetry\footnote{For a recent review of various swampland conjectures see \cite{Palti:2019pca}.}. Since as we have discussed tracking the $U(1)$ gauge symmetry for spin-2 fields is not helpful, it is reasonable to ask whether these more general swampland criterion can be brought to bear on this discussion.

\subsection{Dimensionless Coupling Constants}

An alternative way to try to identify a charge, which is not wedded to a gauge symmetry per se, is to focus on the dimensionless coupling constants that arise in the interaction of the massive spin-2 field and other fields. For instance let us imagine some matter field, which we model as a scalar $\varphi_I$, of mass $m_I$ which is coupled to the massive spin-2 field in manner
\be
- \frac{1}{2} (\partial \varphi_I)^2 - \frac{1}{2} m_I^2 \varphi_I^2+  \frac{1}{M_W}\chi\mn (\partial^{\mu} \varphi_I ) ( \partial^{\nu} \varphi_I) \, ,
\ee
where the mass of the other fields $m_I$ should not be confused with the spin-2 mass $m$.
This coupling does not need to be that of a conserved source a priori. Once again, utilizing the helicity decomposition and integrating by parts, this interaction will generate a dimension 6
interaction
\be
- \frac{2}{M_Wm} A_{\mu} \Box \varphi_I ( \partial^{\mu} \varphi_I) \, .
\ee
We can give this interaction the superficial appearance of a coupling to a current by performing a field redefinition of the matter field
\be
 \varphi_I \rightarrow  \varphi_I - \frac{2}{M_W m} A_{\mu} ( \partial^{\mu} \varphi_I) + \dots\,,
\ee
such that after the redefinition we have an operator of the form
\be
- \frac{2}{M_Wm} A_{\mu} m_I^2  \varphi_I ( \partial^{\mu} \varphi_I) \, .
\ee
This resembles a current interaction $g_I A_\mu J^\mu$, with current $J^\mu=\varphi_I ( \partial^{\mu} \varphi_I)$ and with dimensionless coupling constant
\be
g_I \sim \frac{m_I^2}{M_W m} \, .
\ee
More generally we could have started with mixed interactions in the form
\be
\frac{1}{M_W}\lambda_{IJ}\chi\mn (\partial^{\mu} \varphi_I ) ( \partial^{\nu} \varphi_J)\,,
\ee
and a similar reasoning would give for a current in the form $J_{IJ}^{\nu} = \varphi_I ( \partial^{\nu} \varphi_J)$ and coupling constants
\be
g_{IJ} = \lambda_{IJ} \frac{m_I^2}{M_W m} \, .
\ee
Rather than performing the field redefinition, we may equivalently ask what is the magnitude of the on-shell 3 point function (continued to complex momenta to allow for energy-momenta conservation)  the helicity-1---scalar---scalar vertex.\\

At this point we may run arguments analogous to the swampland conjecture from emergence. In order not to predetermine our answer, we assume that the massive spin 2 states is gapped from the states that define the UV completion so that the latter may be modelled (assuming a weakly coupled UV completion) by an near infinite tower of states whose masses are of the form $m_I = q_I \Lambda_t$ where $q_I$ are dimensionless `charges' for which $q_I \ge 1$. The total number $N$ of such states is assumed to respect the species bound so that the species cutoff $\Lambda_{\rm s}= N \Lambda_t= \mpl/\sqrt{N}$ i.e. $N = (\mpl/\Lambda_t)^{2/3}$.
The coupling constants then take the form $g_I = q_I^2 g$ where $g= \frac{\Lambda_t^2}{M_W m} $.

Following a standard line of reasoning, loop corrections from the heavy states are expected to contribute to the dimensionless coupling constant $g$ from one-loop matter contributions in the amount\footnote{In the usual form of these arguments it is also assumed that the charges are spaced like the mass squareds $g_I \sim q_I g$, i.e. so that $q_I^2$ enters here. This will not substantially change our conclusions, but here we prefer to follow what actually arises from the full spin-2 calculation performed in the next section which fixes the scale based on assumed coupling to UV states. }
\be
\frac{1}{g^2} \sim \frac{1}{g^2_{UV}}+ \sum_I q_I^4  \ln \(\frac{\Lambda_s}{m_I} \) \, .
\ee
This calculation is reflecting the renomalization of the kinetic term for $A_{\mu}$. Assuming that the UV coupling defined at the UV scale $\Lambda_s$ satisfies $1/g_{\rm UV} \sim 0$, i.e. is subdominant to the contribution generated from loops, and in turn assuming that the dimensionless 'charges' are spaced in the manner $q_I \sim n$ then we estimate
\be
\frac{1}{g^2}  \sim N^5 \sim \( \frac{\mpl^2}{\Lambda_t^2}\)^{5/3} \, .
\ee
Putting this together we find
\be
\Lambda_t \sim g^{3/5} \mpl \, ,
\ee
which is similar to the standard magnetic WGC relation, where here $\Lambda_t$ which is the lowest mass in the tower of infinite states is correctly (given our assumptions) identified as the cutoff of the low energy effective theory. Substituting in the scale $g$ we find
\be
\Lambda_t \sim \frac{M_W^3m^3}{\mpl^5} \, ,
\ee
which determines the cutoff of the low energy effective field theory (without the infinite tower of states).
This is an entirely different scale that the conjectured one of \cite{Klaewer:2018yxi}. In particular, since we require $\Lambda_t \gg m$ given our assumptions, we need $M_W^3 m^3 \gg \mpl^5$ which in turn requires $M_W \gg \mpl$.

We thus conclude that, if this reasoning is to be taken seriously, the infinite tower of states must couple to the spin-2 field of mass $m$ more weakly than standard gravitational interactions in order to have the appropriately gapped effective field theory which would contradict the general spirit of the WGC. We will return to this point below in section \ref{reverseordering}.

\subsection{Swampland conjectures in unitary gauge}

The previous argument made use of the swampland conjectures from the idea of emergence, entirely at the level of the helicity-1 mode of the massive spin 2 state. It is helpful to rerun this argument in unitary gauge to give a clear understanding of its implications.  We again begin with a coupling of the form
\be
- \frac{1}{2} (\partial \varphi_I)^2 - \frac{1}{2} m_I^2 \varphi_I^2+  \frac{1}{M_W}\chi\mn (\partial^{\mu} \varphi_I ) ( \partial^{\nu} \varphi_I) \, .
\ee
Integrating out the scalar fields at one-loop and focusing only on the logarithmic divergences (as indeed the above calculation tracks), then we expect corrections to the EFT of the schematic form
\ba
&& \Delta {\cal L} \sim   \frac{a_1}{M_W^2} \chi_{\mu\nu}^2 \sum_I m_I^4  \ln \(\frac{\Lambda_s}{m_I} \)+ \frac{a_2}{M_W^2} \chi^2 \sum_I m_I^4  \ln \(\frac{\Lambda_s}{m_I} \) \\
&& + \frac{b_1}{M_W^2} ( \partial_{\alpha}\chi_{\mu\nu})^2 \sum_I m_I^2  \ln \(\frac{\Lambda_s}{m_I} \)+\frac{b_2}{M_W^2} ( \partial_{\alpha}\chi)^2 \sum_I m_I^2  \ln \(\frac{\Lambda_s}{m_I} \)+\frac{b_3}{M_W^2} ( \partial^{\mu}\chi_{\mu\nu})^2 \sum_I m_I^2  \ln \(\frac{\Lambda_s}{m_I} \) + \dots \nn
\ea
where the coefficients $a_i, b_i$ are of order unity. The $a_i$ coefficients encode precisely the running of the `gauge coupling' $g$  since the kinetic term for the helicity-1 modes comes in unitary gauge from the $\chi_{\mu\nu}^2$ and $\chi^2$ mass terms

With this in mind, the emergence prescription would amount to supposing that the IR mass $m$ is essentially entirely generated by the loop contributions in the form
\be
m^2 \sim \frac{1}{M_W^2} \sum_I m_I^4  \ln \(\frac{\Lambda_s}{m_I} \) \sim \frac{\Lambda_t^4}{M_W^2} \sum_I q_I^4  \ln \(\frac{\Lambda_s}{m_I} \) \sim g^2 \Lambda_t^2\sum_I q_I^4  \ln \(\frac{\Lambda_s}{m_I} \) \sim \frac{\Lambda_t^4 N^5}{M_W^2}\, ,
\ee
which gives the same result as the previous calculation and would again force us to require $M_W\gg \mpl$. By contrast the loop corrections to the coefficient of the kinetic term for the spin-2 field are estimated to be
\be
\Delta Z \sim \frac{1}{M_W^2} \sum_I m_I^2  \ln \(\frac{\Lambda_s}{m_I} \) \sim \frac{\Lambda_t^2}{M_W^2} \sum_I q_I^2  \ln \(\frac{\Lambda_s}{m_I} \) \sim \frac{\Lambda_t^2}{M_W^2} N^3 \sim  \frac{\mpl^2}{M_W^2} \ll 1\, .
\ee
This implies that the kinetic term for the massive spin-2 field cannot itself be viewed as emergent, assuming the validity of the aforementioned scalings.

\subsection{$M_W \gg \mpl$?}

\label{reverseordering}

At first sight, the conclusion that $M_W \gg \mpl$ goes against the WGC since it implies that any massive spin-2 field couples more weakly than the standard (massless) graviton. In practice however, it just points to the over simplification of these arguments. To see why, let us consider a theory in which the normally massless graviton acquires a mass, but most species of matter are coupled covariantly to the metric so that $M_W=\mpl$. In other words the action for our matter fields is
\be
\sqrt{\eta_{\mu\nu}+ \mpl^{-1}\chi_{\mu\nu}}\left[ - \frac{1}{2} ((\eta+\mpl^{-1} \chi)^{-1})^{\mu\nu}\partial_{\mu} \varphi_I \partial_{\nu} \varphi_I - \frac{1}{2} m_I^2 \varphi_I^2\right] \, .
\ee
Expanding this to first order in $\mpl^{-1}$ gives exactly the type of interaction considered previously. Since this coupling is covariant, integrating out the matter fields will only generate covariant interactions \cite{deRham:2013qqa}. The leading non-derivative interactions come entirely in the form of a cosmological constant, as
\be
\Delta {\cal L} \sim -c \sum_I m_I^4 \  \sqrt{\eta_{\mu\nu}+ \mpl^{-1}\chi_{\mu\nu}}\ \ln \(\frac{\Lambda_s}{m_I}\)\,,
\ee
which must be cancelled by a counterterm to remove the tadpole contribution. When this is done the net result is that the loop corrections to the mass are zero at one-loop
\be
\Delta m^2 \sim 0 \, .
\ee
This is of course diffeomorphism invariance at work. Since matter couples covariantly, we cannot generate a mass term which itself breaks diffeomorphisms. What is happening is that in this case the naively order unity coefficients $a_1$ and $a_2$ actually vanish, at least at one-loop order, after the tadpole has been cancelled. The correct conclusion of the previous estimate should have been that
\be
m^2 \sim \frac{|a_i| \Lambda_t^4 N^5}{M_W^2}\,,
\ee
which given the requirement $\Lambda_t \gg m$ enforces
\be
|a_i |\ll \( \frac{m^2 M_W^3}{\mpl^5}\)^{2/3} \, .
\ee
Thus we are allowed to preserve the naive WGC expectation $\mpl \ge M_W$ provided the breaking of diffeomorphism invariance in the coupling of the spin-2 particle to other species is suppressed by powers of $M_W/\mpl$, which was shown to be  natural \cite{deRham:2013qqa,deRham:2014naa}. Indeed, this is a tuning we would expect to be natural in any mechanism whereby the usual massless graviton acquires a mass, such as in the mechanism proposed in \cite{Bachas:2011xa,Bachas:2017rch,Bachas:2018zmb,Bachas:2019rfq}.

\section{Example interacting spin-2 theories}
\label{sec:examples}

\subsection{Kaluza-Klein theory}

In {\cite{Klaewer:2018yxi} the bound \eqref{eq:conjecture}} is applied to Kaluza-Klein (KK) theory and it is argued that this gives evidence of its validity. In the original KK theory for 5 dimensions compactified on a circle of radius $L$, there is an infinite tower of massive spin-2 states. However, in addition to the massive spin-2 field and therefore {\bf in addition to the helicity-1 mode $A_\mu$} introduced in \eqref{eq:helicitydecomp}, crucially there are 5 other zero modes, the massless graviton $g\mn$, a gravi-photon $B_\mu$ and a dilaton/radion $\phi$. The gravi-photon $B_\mu$ is a genuine massless $U(1)$ gauge field (that we shall designate as $U(1)_B$ to distinguish it from the `fake' $U(1)_A$). The global symmetry for $U(1)_B$ is associated with translations in the extra dimension and has nothing to do with the helicity-1 modes $A_\mu$ of the massive spin-2 field tower. The four-dimensional $U(1)_B$ gauge symmetry is just a remnant of the higher dimensional symmetry. Crucially, in this case there {\bf does} exist a meaningful global limit of this $U(1)_B$ gauge field associated with the gravi-photon $B_\mu$. {However, this} genuine $U(1)_B$ has nothing to do with the `fake' $U(1)_A$ that arises in the massive states { and therefore the situation in a Kaluza-Klein theory is dramatically different to that of a single massive spin-2}. This $U(1)_B$ is a {completely} separate $U(1)$ gauge field under which the massive KK modes themselves are charged.
Ignoring the dilaton/radion (which is largely irrelevant to the story) we denote the 5 dimensional metric as
\be
\d s^2 = \left(\d y+\frac{B_{\mu}(x)}{M_4} \d x^{\mu}\right)^2 +  \left( g_{\mu\nu}(x)+ \frac{2}{M_4} \sum_{n\neq 0} \chi^n_{\mu\nu}(x) e^{2 \pi i n y/L} \right) \d x^{\mu} \d x^{\nu}\,,
\ee
where $\chi^n_{\mu\nu}={\chi^{-n}_{\mu\nu}}^*$ are the massive KK modes. As is well-known, this form of the five dimensional metric exhibits a $U(1)_B$ symmetry which descends from the $y$ diffeomorphism symmetry. Specifically, the form of the metric is invariant under the $U(1)_B$ transformation
\ba
\label{eq:translation1}
&& y \rightarrow y -\frac{1}{M_4}\chi(x) \\
&& B_{\mu}(x) \rightarrow B_{\mu}(x) + \partial_{\mu} \chi(x) \\
&& \chi^n_{\mu\nu}(x) \rightarrow \chi^n_{\mu\nu}(x) e^{ 2 \pi i n \chi/(LM_4)}  \, ,
\label{eq:translation3}
\ea
under which it is clear that the KK modes transform as charged fields with charges $g_n= 2\pi n/(M_4L)$. The global limit of this symmetry $\chi(x) = \text{constant}$ is translation symmetry in the extra dimension. We emphasize that in writing (\ref{eq:translation1}--\ref{eq:translation3}), we have made the diffeomorphism gauge choice to work in unitary gauge for each $\chi^n\mn$, reason why the helicity-1 modes $B_\mu^n$ do not appear. But we could have very well reintroduced the $\ldiff \times U(1)_A$ symmetries by introducing a set of  \stu fields $B_\mu^n$ and $\pi^n$ for each $\chi\mn^n$ and we would have then also got multiple copies of the symmetry (\ref{eq:diff1}--\ref{eq:diff3}) that includes multiple copies of $U(1)_A$. Clearly, as we shall see below none of those $U(1)_A$ play any role for how the WGC applies to Kaluza-Klein theories. \\

Expanding the five-dimensional Einstein-Hilbert action
\be
S = \frac{M_5^3}{2}  \int_0^L \d y \int \d^4 x \sqrt{-g_5} R[g_5]\,,
\ee
with $M_5^3=M_4^2L$ and integrating over the size $L$ of the extra dimension,
to quadratic order in the KK modes we obtain schematically
\ba
S= \int \d^4 x \sqrt{-g} &\Bigg(&  \frac{M_4^2}{2}   R- \frac{1}{4} F_{\mu\nu}^2  + \sum_{n=1}^{\infty} \frac{1}{2} \chi_n^{\mu \nu}{}^* {{\cal E}}\left[\partial_{\mu}- i \frac{n}{M_4L}B\right] \chi^n_{\mu\nu} \\
&&-\frac{1}{2}\sum_{n=1}^\infty \left(\frac{2 \pi n}{L}\right)^2 (\chi^{n*}_{ \mu\nu}\chi_n^{\mu\nu}-|\chi_n|^2)+   \dots \Bigg)\notag
\ea
where ${{\cal E}}[\partial_{\mu}-  \frac{in}{M_4L}B] $ is a $U(1)_B$ gauge covariant version of the Lichnerowitz operator  and $F_{\mu\nu}$ is the strength tensor gravi-photon gauge field $B_\mu$. The gauge coupling is $g= 2 \pi /(M_4L)= 2\pi/(M_5L)^{3/2}$ and the lightest massive graviton state has mass $m=2\pi/L$. In this case the WGC applied to the lightest mass state which is charged under the $U(1)_B$ gives a cutoff
\be
\Lambda \sim g M_4 \sim \frac{1}{M_4 L} M_4 \sim \frac{1}{L} \,,
\ee
which is exactly the result we expect since having integrated out the extra dimension, the cutoff of the EFT for the lightest massive spin-2 state will be at the scale of the next lightest massive spin-2 state. However, this does not provide any evidence for the conjecture of {\cite{Klaewer:2018yxi}} since this argument only applies because there is a separate $U(1)_B$ gauge field under which the massive spin-2 states are charged, whereas the claim of  {\cite{Klaewer:2018yxi}} is that this bound should apply for uncharged spin-2 particles. Hence, the claimed evidence in \cite{Klaewer:2018yxi} is non-existent.\\

Five-dimensional  diffeomorphisms are realized in four dimensions as gauge version of a Kac-Moody-like algebra \cite{Dolan:1983aa}. At the quadratic level, the \stu fields are made manifest by writing the five dimensional metric perturbation in the form
\ba
\d s^2 &=& \left( 1+4 \sum_{n}  e^{i 2\pi n y/L}  \frac{\pi_n(x)}{M_4} \right)\d y^2+4 \sum_{n} e^{i 2\pi n y/L} \frac{A^n_{\mu}(x) }{M_4} \d x^{\mu}  \d y \\
& + &  \left( \eta_{\mu\nu}+ \frac{2}{M_4} \sum_{n\neq 0} \chi^n_{\mu\nu}(x) e^{2 \pi i n y/L} \right) \d x^{\mu} \d x^{\nu}\,,\nn
\ea
for which $A^0_{\mu}=B_{\mu}$. Linear 5D diffeomorphisms are realized as
\ba
&& \pi_n \rightarrow \pi_n +  m_n \chi_n  \\
&& A^n_{\mu} \rightarrow A^n_{\mu} + i ( m_n \chi^n_{\mu}-\partial_{\mu} \chi_n) \\
&& \chi^n_{\mu\nu}\rightarrow \chi^n_{\mu\nu} + \partial_{\mu} \chi^n_{\nu}+ \partial_{\nu} \chi^n_{\mu}\,,
\ea
with $m_n = 2 \pi n/L$ the KK masses which for $m_n \neq 0$ we recognize as the massive spin-2 \stu gauge symmetry. The generator associated with $A^n_{\mu}$  is $Q_n=i e^{i 2\pi n y/L} \partial_y$ and it is these $Q_n$ that in the linear theory are the $U(1)_A$'s associated with the helicity-1 modes of the massive KK modes. Nonlinear however these symmetries are non-abelian.
Individual KK modes are not eigenstates of $Q_n$ for $n \neq 0$ but they are eigenstates of $Q_0$. The $Q_{n\ne 0}$ symmetries are non-abelian are nonlinearly realized, and the vacuum is not invariant under them.

\paragraph{Modding out the Circle:} At this point, one may be tempted to project out the genuine vector field $B_\mu$ by modding out the extra dimension\footnote{We would like to thank the authors of~\cite{Klaewer:2018yxi} for suggesting this interesting approach.} by a $\mathbb{Z}_2$. Such a procedure would indeed remove the massless boson $B_\mu$, and superficially removes the associated global charge by combining complex $n$ and $-n$ modes into real uncharged fields. However the scale of interactions is set by the unprojected theory which itself contains no mass gap in order to talk about a separate low energy effective theory for a finite number of massive spin-2 states. To summarize, what this KK setup provides is an explicit example the standard WGC applied for spin-1 fields. This is well known example and it provides no connections with the actual claims of a WGC for massive spin-2 fields.

\subsection{Bigravity and Massive Gravity}

An interesting test case of these theories are the bigravity and massive gravity theories. Since massive gravity arises as a consistent scaling limit of bigravity we may first consider the former. Consider a theory with two different metrics $g$ and $f$, with associated Planck scales $M_g$ and $M_f$ coupled together by some mass term whose quadratic form takes the Fierz-Pauli structure. For instance an explicit nonlinear form known to introduce no new degrees of freedom classically is \cite{deRham:2010kj}
\ba
U(g^{-1}f)= 4 \( \({\rm Tr} \left[\sqrt{g^{-1}f}-1\right]\)^2- {\rm Tr}\left[(\sqrt{g^{-1}f}-1)^2\right] \) \,.
\ea
Hence, the bigravity Lagrangian is given by \cite{Hassan:2011zd}
\be
{\cal L} = \frac{1}{2}\sqrt{-g} M_g^2 R[g] +\frac{1}{2} \sqrt{-f} M_f^2 R[f] - \frac{1}{8} \sqrt{-g} m^2 \tilde M^2 U(g^{-1}f)+ \dots
\ee
Interactions between the two metrics breaks two copies of diffeomorphisms down to a single copy and the metric eigenstates $g$ and $f$ include an admixture of both the mass eigenstates, the massless graviton metric\footnote{We stress that $g^0_{\mu\nu}$ is not a fixed background metric but the dynamical metric of the massless graviton $h_{\mu\nu}$, i.e. $g^0_{\mu\nu}=\eta_{\mu \nu}+ h_{\mu\nu}/\mpl$.} $g^0_{\mu\nu}$ and the massive graviton $\chi_{\mu\nu}$. The diagonalization from metric to mass eigenstates is achieved as follows: we define \cite{Hassan:2012wr}
\ba
g_{\mu\nu}= g^0_{\mu\nu} + \frac{2 \tilde M}{M_g^2}\chi_{\mu\nu} \\
f_{\mu\nu}= g^0_{\mu\nu} - \frac{2 \tilde M}{M_f^2}\chi_{\mu\nu} \, ,
\ea
where the mass scale $\tilde M$ is defined as
\be
\frac{1}{\tilde M^2} = \frac{1}{M_g^2}+ \frac{1}{M_f^2} \, .
\ee
From this decomposition it is clear that if $M_f \gg M_p$ then the $f$ metric is 'mostly massless' and the $g$ metric is 'mostly massive', and if $M_f \ll M_p$ this reverses.

Then expanding to quadratic order in $\chi$ we recover the action for a massive spin-2 field covariantly coupled to a dynamical metric $g_0$ for which $\mpl^2 = M_g^2+M_f^2$,
\be
{\cal L} = \sqrt{-g_0}\left[\frac{1}{2}  \mpl^2 R[g_0] +\frac{1}{2} \chi^{\mu\nu} {\cal E}[g_0] \chi_{\mu\nu} - \frac{1}{2}m^2 (\chi^{\mu\nu} \chi_{\mu\nu} -\chi^2)\right]+ \dots\,,
\ee
and the self interactions of $\chi$ are at the scale
\ba
M_W= {\rm Min}\(\frac{M_g^2}{\tilde M}, \frac{M_f^2}{\tilde M}\)= {\rm Min}\(\frac{M_g}{M_f}, \frac{M_f}{M_g}\) \mpl \ \le  \ \mpl\,.
\ea
Crucially, we note that bigravity models always satisfy the WGC in the sense that $M_W \le \mpl$. For instance, in the usual context of massive bigravity models we implicity assume a hierarchy of Planck masses, e.g. $M_f  \gg M_g$. In this hierarchy $\mpl \sim M_f$, $\tilde M \sim M_g$ and {$M_W \sim M_g$} and so we automatically have $M_W \ll \mpl$.
At the level of the Lagrangian this arises because every metric contains a part which is the massless graviton, and these parts sum up positively, whereas the massive parts add and subtract.

The conjecture of \ref{eq:conjecture} would imply an upper bound on the cutoff of
\be
\Lambda \le \frac{m M_f}{M_g} \, .
\ee
Given the assumed hierarchy $M_f \gg M_g$, {even if the conjecture were true} this bound is not necessarily problematic since $\Lambda \gg m$ and so it still makes sense to talk about the effective theory of the massive spin-2 state coupled to gravity. Massive gravity is obtained from bigravity in the limit $M_f \rightarrow \infty$ keeping $m$ and $M_g$ fixed.

Thus, in this limit the conjectured upper bound in {\cite{Klaewer:2018yxi} actually} asymptotes to infinity. In {other words} massive gravity automatically {would satisfy} the WGC in this form.

\subsection{Higher derivative gravity}

Let us now consider the limit $M_f \ll M_g$ so that $\mpl \approx M_g$ and $g$ is the mostly massless metric. At energies well below the mass $m$ of the massive spin-2 state we may integrate it out resulting in an covariant theory with an infinite number of derivatives whose regime of convergence is $R \ll m^2$, $\nabla \ll m$. An explicit example of this is \cite{Gording:2018not}. Although the theory with an infinite number of higher derivatives is formally ghost-free, since the regime of convergence of the derivative expansion is  $\nabla \ll m$ we must treat it as an EFT with a cutoff $\Lambda \sim m$.
Interestingly in 3 dimensions this procedure may be performed giving rise to only a finite number of higher derivative terms \cite{deRham:2011ca,Paulos:2012xe} leading to new massive gravity \cite{Bergshoeff:2009hq}. As noted in \cite{deRham:2011ca} the new massive gravity action can be obtained from integrating out a second spin-2 field in the form
\be
S = \frac{M_3}{2} \int \d^3 x \sqrt{-g} \left[ -R-\chi^{\mu\nu} G_{\mu\nu}- \frac{1}{4} m^2 (\chi_{\mu\nu}^2-\chi^2)\right]\,,
\ee
which exhibits the Fierz-Pauli structure. This in turn can be obtained as a scaling limit of bigravity \cite{Paulos:2012xe}.

In general dimensions, for ghost free bigravity models with specific coefficients the leading terms in a derivative expansion take the form
\be
{\cal L} = \frac{1}{2}\mpl^2 R[g] + \frac{1}{2 g_W^2} W_{\mu\nu\rho\sigma}W^{\mu\nu\rho\sigma}+ \dots
\ee
 and we are using the fact that in this limit $M_g \approx \mpl$. Taken on its own this truncated Einstein--Weyl--squared action propagates a massless spin-2 field and a ghostly massive spin-2 field \cite{Stelle:1976gc,Stelle:1977ry}. The mass of the ghostly spin-2 mode is $m_{\rm ghost}\sim g_W \mpl$. In \cite{Klaewer:2018yxi}, the conjecture \eqref{eq:conjecture} is then applied to this ghostly massive spin-2 state to give
\be
\Lambda_m \sim \frac{m_{\rm ghost} \mpl}{M_W} =\frac{g_W \mpl^2}{M_W} =  \frac{g_W M_g^2}{M_f} \, ,
\ee
thus defining a WGC for the Einstein--Weyl--squared theory. This identification is simply inconsistent since the {\bf massive spin-2 ghost state that arises in the truncated EFT is unrelated to the physical massive spin-2 state in the UV theory} and in particular there is no reason to identify $m$ with $m_{\rm ghost}$ as we show below.\\

\paragraph{Cutoff of truncated expansion vs mass of integrated mode:} To illustrate why the massive spin-2 ghost in the truncated EFT cannot be directly identified with physical massive spin-2 state in the UV theory that has been integrated out, let us consider a simple example of a weakly coupled UV completion of a scalar field for which the momentum space propagator exhibits a K\"allen-Lehman spectral representation with only poles (weakly coupled) and positive residues
\be
G(k)= \frac{1}{1+\lambda^2 } \left[ \frac{1}{m_0^2+k^2}+ \frac{\lambda^2}{M^2+k^2} \right] = \frac{\frac{M^2+\lambda^2 m_0^2}{1+\lambda^2}+k^2}{(m_0^2+k^2)(M^2+k^2)} \, .
\ee
The representation is chosen so that $\lambda$ can take any value without spoiling unitarity and so that $G(k^2)\sim 1/k^2$ at high energies \cite{Schwinger:1959wya}. Assuming $M \gg m_0$, at energies $k^2 \ll M^2$ we can expand the inverse propagator in inverse powers of $M^2$ to derive the low energy EFT expansion
\be
G(k)^{-1}= (1+ \lambda^2) (k^2+m_0^2) \( 1- \frac{\lambda^2}{M^2} (k^2+m_0^2)\) + {\cal O}(1/M^4) \, .
\ee
Truncating this expansion at this order then we would infer the existence of the original light pole $m_0$, and a ghostly pole whose mass is $m_{\rm ghost}^2 = m_0^2 -M^2/\lambda^2$
\ba
\hspace{-1.5cm}G(k)_{\rm truncated} = \frac{1}{(1+ \lambda^2) (k^2+m_0^2) \( 1- \frac{\lambda^2}{M^2} (k^2+m_0^2)\) } =\frac{1}{1+\lambda^2 }\frac{1}{m_0^2+k^2}- \frac{1}{1+\lambda^2 }\frac{1}{k^2+m^2_{\rm ghost}} \, .\hspace{-1cm}
\ea
We see that the mass of the ghost in the truncated theory bears no resemblance to additional massive state in the UV completion. Indeed $m_{\rm ghost}^2$ is tachyonic and tends to $-\infty$ as $\lambda \rightarrow 0$ whereas the actual mass $M^2$ remains finite.
A similar story holds in the spin-2 case, just following from the spin-2 K\"allen-Lehman spectral representation. Thus, we see that even if the spin-2 conjecture of \cite{Klaewer:2018yxi} were correct, its application to the ghostly massive spin-2 state would be unjustified since the mass of the actual physical spin-2 state bears no resemblance to that of the spin-2 ghost.\\

The actual cutoff of the truncated EFT, or alternatively the regime of convergence for the derivative expansion is, clearly $\Lambda = {\rm Min}(M,|m_{\rm ghost}|)$. If $\lambda < 1$ then $\Lambda=M$ as we would expect from having integrated out the massive state. If $\lambda>1$ then the cutoff is lower than $M$ since the physical massive spin-2 particle becomes more dominant and its negligence at low energies becomes increasingly invalid, despite its large mass.

Indeed more generally, given any truncated higher derivative gravity theory, there will be additional ghostly-spin-2 modes. The ghosts are not a problem since they signify only that the EFT is breaking down (see for example \cite{deRham:2014fha}), and so the standard EFT cutoff is at most the mass of the lightest such ghost but may even be less as in the above example with $\lambda>1$.

\subsection{String theory example}

In string theory, spin-2 states can arise in the infinite tower of excitation of string modes. Recently an interesting string theory UV completion for the supergravity version of the Einstein--Weyl--squared theory was proposed in Ref.~\cite{Ferrara:2018wlb}. The string set up is 10 dimensional string theory compactified on $R^{1,3} \times T^6$ with a stack of $N$ $D3$-branes whose open string sector includes in addition to the Yang-Mills sector, massive spin-2 excitations as well as an infinite tower of other states. Ref.~\cite{Ferrara:2018wlb} proposed the interesting claim that these stringy massive spin-2 states could be identified with the ghostly massive spin-2 states that arise in the supergravity version of Einstein--Weyl--squared. However, as we have seen above from the K\"allen-Lehman spectral representation arguments such an identification cannot be trusted. Both the physical UV spin-2 state and the ghostly one may sit in the same supermultiplet, but this does not identify them. The Einstein--Weyl--squared supergravity theory should be correctly interpreted as an EFT for which the Weyl--squared term should only ever be treated perturbatively, i.e. for which the ghost pole never actually arises.

In order to correctly identify the low energy EFT associated with the string setup, it is necessary to show that the associated Weyl--squared terms correctly reproduce the leading corrections to the scattering amplitude for the {\bf massless} graviton and the other massless states that propagate within the low energy EFT. This cannot be achieved by trying to connect unphysical poles with physical poles in the truncated theory. Indeed we would expect all the low excited states to contribute since for example the 2--2 scattering amplitude for massless gravitons can receive exchange contributions for particles of any intermediate spin.

Once again we can illustrate this point with a simple example. Suppose we want to compute the 2--2 scattering amplitude of a massless scalar, e.g. dilaton, in the low energy effective theory. If the UV completion is weakly coupled, as in the case of string theory with $g_s \ll 1$, then the UV 2--2 scattering amplitude $A_{\rm UV}(s,t)$ will take the form of a sum of poles whose residues are determined by the coupling $g_{I,J}$ of the scalar to the excited string states of mass $m_{I,J}$ and spin $J$\footnote{For a recent application of S-matrix formulae of this type see \cite{Caron-Huot:2016icg}.}:
\be
A_{\rm UV}(s,t) = a(t) + b(t) s + \frac{\lambda_0^2}{(-s)}+\frac{\lambda_0^2}{(-u)}+ \frac{s^2}{\pi} \sum_{I=1}^{\infty} \sum_{J=0}^{\infty} \frac{g_{I,J}^2 P_J\left (1+\frac{2t}{m^2_{I,J}}\right)}{m^4_{I,J}(m^2_{I,J}-s)}+ \frac{u^2}{\pi} \sum_{I=1}^{\infty} \sum_{J=0}^{\infty} \frac{g_{I,J}^2 P_J\left (1+\frac{2t}{m^2_{I,J}}\right)}{m^4_{I,J}(m^2_{I,J}-u)} \, .
\ee
An intermediate spin-2 particle shows up in this expansion via the Legendre Polynomial $P_2(1+2t/m^2_{I,J})$.
By contrast, the low energy effective theory will simply give an expansion in powers of $s$ and $t$
\be
A_{\rm EFT}(s,t) =  \frac{\lambda_0^2}{(-s)} +\frac{\lambda_0^2}{(-u)} +\frac{\lambda_0^2}{(-t)} + \sum_{n,\ell \ge 0} c_{n,\ell} s^n t^\ell \, ,
\ee
which can be matched to $A_{\rm UV}(s,t)$ by expanding in inverse powers of $m^2_{I,J}$.
Since every intermediate spin $J$ contributes a term at every power in $t$ up to $t^J$ already from the $s$-channel, a given coefficient $c_{n,\ell}$ will receive contributions from all spins $J \ge \ell$ from the $s$-channel and all spins from the $u$-channel. A similar argument holds for scattering of external states of arbitrary spin.

In the present context, this means that in the actual low energy EFT related to the string set up, the coefficient of the Weyl--squared term will receive contributions from all excited states of spins $J\ge 2$. Thus, purely at the scattering amplitude level, it is impossible to identify the Weyl--squared term as having come from the massive spin-2 excitation alone.

Now, let us turn to the application of the conjecture \eqref{eq:conjecture}  to the string theory model as proposed in \cite{Klaewer:2018yxi}. Since this model is itself UV complete, we do not expect any problem. However, we can ask what would happen if we integrated all excited string modes other than the massless ones and the massive spin-2 supermultiplet. Since the spectrum of excited modes takes the form $m_n \sim \sqrt{n}\, m$ where $m=g_sM_s$, the cutoff of the resulting effective theory will clearly be at most of order $m$. The conjecture \eqref{eq:conjecture}  of \cite{Klaewer:2018yxi} supposes that the cutoff for the lightest spin-2 state is $\Lambda \sim m \mpl/M_W$ and if we naively identify $\mpl=M_W$ then the conjecture is $\Lambda \sim m$. In this case the conjecture is in a sense correct because the known cutoff is at most $m$, however its validity here is more of an accident since it relies on an identification that cannot be trusted. Furthermore the identification of $M_W$ with $\mpl$ cannot be trusted since the open string modes may couple quite differently than the closed string modes. Crucially though the fake \stu $U(1)_A$ symmetry plays no role here, unlike the true $U(1)_B$  symmetry in the Kaluza-Klein example. Rather what this string model illustrates is an example of a standard weakly coupled UV completion in which the spin-2 state arises in an infinite ungapped tower.

\subsection{Nonlinear massive gravity}

Massive gravity theories are obtained from bigravity by taking the scaling limit $M_f \rightarrow \infty$. This makes the massless graviton infinitely weak, naively consistent with the WGC, decoupling its fluctuations. The only remaining relevant part of the metric $f$ is to act as a fixed reference metric, which may for example be taken to be Minkowski spacetime to preserve a global Poincar\'e symmetry. In nonlinear theories of massive gravity, for which the symmetry spontaneously broken is the nonlinear Diffeomorphism group, the \stu  fields $A_{\mu}$ and $\pi$ arise as part of a set of four scalar fields
\be
\phi^a = x^a + \frac{1}{m M_g} A^a + \frac{1}{m^2 M_g}\partial^a \pi \, .
\ee
In this case it is clear that the $U(1)_A$ symmetry for which $A_{\mu}$ is the gauge field is entirely an artifact of the decoupling limit since $\phi^a$ is not a vector under the diffeomorphism group but a set of 4 scalars. Rather they are vectors under an additional global Poincar\'e group which acts to leave invariant the reference metric $f_{\mu\nu} = \eta_{ab} \partial_{\mu} \phi^a \partial_{\nu} \phi^b$. Furthermore $\pi$ is not a diffeomorphism scalar, and its action is not covariant in the usual sense (e.g. setting $A^a$ to zero does not give rise to a covariant action for $\pi$).

There is a profound consequence of this which is that the current $J^{\mu}$ for the vector field receives no direct contribution from matter in the usual theory in which matter couples minimally to the metric $g$ (see \cite{deRham:2014naa,Yamashita:2014fga,deRham:2014fha} for discussions of allowed couplings). In other words, the breaking of diffeomorphisms by the mass term does not in itself lead to a contribution to $J^{\mu}$ directly from matter. Rather it comes indirectly through the mass term which is itself a nonlinear function of the \stu fields.

In reference {\cite{Klaewer:2018yxi}} a stronger spin-2 conjecture is claimed for massive gravity that $\Lambda_m \sim m$. This is on the grounds that `all the evidence presented holds equally of the lightest spin-2 state is actually not massless'. Here the idea is to rerun all the arguments leading to the support of the WGC for a massive spin-2 field. However all these arguments fall apart in this case. For instance, the nature of black holes in a theory of pure massive gravity is completely different (for a recent discussion see \cite{Rosen:2017dvn,Rosen:2018lki}). It is already true that the uniqueness theorems no longer apply, and that there are no asymptotically flat static solutions. Given this, the usual arguments pertaining to black holes carrying global charges cannot be carried through. {Furthermore}, since the {naive} unitarity cutoff of these theories $(m^2 M_g)^{1/3}$ is typically smaller than the inverse Schwarzschild radius it is not even clear we can talk about black holes in the regime of validity of this effective theory (to do so implicitly assumes the Vainshtein mechanism redresses the cutoff scale).

Invoking a proper WGC argument on massive spin-2 fields would require first identifying a global symmetry that is in conflict with coupling to massless gravity. One possible direction is to work with the global Poincar\'e symmetry. We may for instance worry that massive gravity has a conserved global Poincar\'e current associated with transformations $\phi^a \rightarrow \phi^a + c^a + \Lambda^{a}{}_b \phi^b$. Coupling this to a massless graviton however immediately turns this global symmetry into a local one, i.e. diffeomorphism invariance. This is very different from an internal gauge symmetry which exist in a global limit coupled to gravity. Indeed another important distinction is that the global Poincar\'e symmetry is always non-linearly realized, and there are no states in the free theory charged under this symmetry.
It seems unlikely that an argument along these lines would conclude anything other than $M_W < \mpl$.

\subsection{A UV completion of Massive gravity and Bigravity}

The absence of vDVZ discontinuity on Anti-de Sitter (AdS) indicates that a UV completion of massive gravity and bigravity is likely to be more natural on AdS \cite{Higuchi:1986py,Kogan:2000uy,deRham:2015ijs,deRham:2016plk,deRham:2018svs} and higher dimensional embeddings have been attempted for more than two decades, either from integrating out conformal fields in AdS with transparent boundary conditions \cite{Porrati:2000cp,Porrati:2001gx,Porrati:2001db,Porrati:2002cp,Porrati:2003sa}, or by embedding an AdS brane in an AdS Karch-Randall extra dimension \cite{Karch:2000gx,Karch:2001jb}, the two pictures being holographically dual to each other.

In trying to construct string theory realizations of the Karch-Randall scenario, in general no clear separation of scales between the graviton mass and the AdS curvature could be achieved, hence potentially leading to a construction that would once again carry a cutoff at the scale of the graviton mass. However,  a very promising new approach has been proposed recently involving  highly-curved Janus throats that connects different AdS regions of space \cite{Bachas:2018zmb}. In this UV realization, the graviton mass can be made parametrically smaller than the AdS curvature and the cutoff of the theory, \cite{Bachas:2019rfq}.  An explicit realization for bi-gravity is also discussed in \cite{Bachas:2017rch}. As is usual, the cutoff of the low energy effective theory is expected to go to zero in the limit $m\rightarrow 0$. More precisely ref. \cite{Bachas:2019rfq} conjectures that $m=0$ is at an infinite distance in moduli space, and approaching this point brings down a tower of spin $\ge 2$ excitations with mass spacings vanishing at least as $\Lambda_*$ where $\Lambda_*^{D+2}=m^{2} \mpl^{(D-2)}l_{AdS}^{-2}$, where $l_{AdS}$ is the AdS length scale in $D$ dimensions. This is the standard scale expected on anti-de Sitter spacetime (see for example \cite{deRham:2015ijs,deRham:2016plk,deRham:2018svs}) and the construction of \cite{Bachas:2017rch,Bachas:2018zmb} is consistent with this conjecture. Assuming this construction is indeed completely under control, it represents an explicit UV completion of massive gravity/bigravity in anti-de Sitter spacetime, and in turn an explicit counter example to the arguments of \cite{Klaewer:2018yxi}.

\section{Positivity Bounds and Asymptotic (sub)luminality}
\label{sec:Positivity}
We have shown that the arguments presented in \cite{Klaewer:2018yxi} do not meaningfully imply any type of weak gravity conjecture on massive spin-2 fields and cannot be used to impose an upper bound on the cutoff of a massive spin-2 EFT. However this is not to say that such bounds do not exist nor to say that a proper WGC could not in principle be adequately derived for massive spin-2 field.  First we note that besides WGC considerations, {\it standard EFT weak coupling} arguments already provide by themselves robust bounds on the cutoff of massive spin-2 fields. 

When dealing with EFTs it is standard to associate the strong coupling scale $\Lambda_{*}$, i.e. the scale at which perturbative unitarity is broken, with the cutoff $\Lambda_{\rm cutoff}$ of the theory. This is because in {\it weakly coupled} theories new physics should enter at or below the scale $\Lambda_*$ to restore unitarity. This strongly relies on the assumption of a `weakly coupled UV completion' and in principle unitarity could be preserved in a strongly coupled UV completion without the need for new states \cite{Aydemir:2012nz,deRham:2013qqa,deRham:2014wfa}. However if we assume a weakly coupled UV completion, then the cutoff, i.e. the scale at which new physics comes in, should be comparable or lower than the strong coupling scale $\Lambda_{\rm cutoff}<\Lambda_*$. For local and Lorentz-invariant  massive spin-2 fields (including but not restricted to massive gravity and bigravity) this implies that, the cutoff should be at most $\Lambda_{\rm cutoff} \lesssim (m^2 M_W)^{1/3}$ (unless classical Vainshtein redressing is invoked). A similar bound also exist for charged massive spin-2 fields \cite{Porrati:2008ha,deRham:2014tga}.

A recently developed powerful constraining tool for the low-energy EFTs is the framework of positivity bounds which are independent and complementary to the Swampland and WGC conjectures. These bounds arise by exploiting the powerful implications of S-matrix unitarity and analyticity to infer important properties of the low-energy EFT in order for it to have a Lorentz invariant local and unitary UV completion. The resulting positivity bounds restrict the signs of the Wilson coefficients in the effective Lagrangian and represents crucial theoretical constraints of the parameter space of the allowed EFT\footnote{For a recent application to EFT corrections to the Standard Model Lagrangian see \cite{Zhang:2018shp}.}.

A crucial precondition of this framework is the presence of a mass gap which can have important consequences \cite{deRham:2017imi}. This theoretical testing ground can also be directly applied to massive spin-2 fields \cite{Bonifacio:2016wcb} and massive gravity \cite{Cheung:2016yqr,Bellazzini:2017fep,deRham:2017xox,deRham:2018qqo}, placing restrictions on the allowed parameter space where the theory could admit local, unitary and Lorentz invariant UV completion.

In its original formulation \cite{Adams:2006sv} the positivity bounds successfully constrain the sign or allowed regions for operators \cite{Cheung:2016yqr} but provide little bounds on the actual cutoff\footnote{Note however that for some EFTs the positivity bounds are so powerful at eliminating entire classes of interactions that the resulting cutoff can be affected \cite{deRham:2018qqo}.}. However an extension of these bounds (involving scattering amplitudes beyond the forward limit) as presented in \cite{deRham:2017avq,deRham:2017zjm} directly involve the cutoff and can be used in certain cases to place an upper bound on the cutoff of the low-energy effective field theory.

In their original formulation, the positivity bounds use positivity of the imaginary part of the scattering amplitude throughout the physical region (at energies above $2m$, where $m$ is the mass of the field considered in the low-energy EFT). However assuming weak coupling, the scattering amplitude can be consistently computed up to the cutoff of the EFT and can be subtracted out from the positivity bounds leading to much more stringent or `improved' positivity bounds \cite{Bellazzini:2016xrt,deRham:2017imi,deRham:2017xox} that directly involve an upper bound of the cutoff of the theory. Within the context of massive Galileons (which are typical interactions that arise in self-interacting massive spin-2 fields),  the bounds on the cutoff was derived in \cite{deRham:2017xox}. For massive gravity, initial attempts in applying those bounds were provided\footnote{Note that the paper \cite{Bellazzini:2017fep} contains confusions about the application of EFTs that were clarified in \cite{deRham:2017xox} for which the correct implications of massive gravity were discussed.} in \cite{Bellazzini:2017fep}. The proper account for the meaning of weak coupling in massive gravity and the implications of the improved positivity bounds was then given in \cite{deRham:2017xox}, where it was shown that the improved positivity bounds are satisfied so long as the coupling constant $g_*$ is
\begin{eqnarray}
\label{eq:g*}
g_* \lesssim \left(\frac{m}{M_W}\right)^{1/4}\ll 1\,,
\end{eqnarray}
where $M_W$ is the typical scale of the spin-2 self-interactions and we assume a standard strong coupling at the scale $\Lambda_3=(m^2 M_W)^{1/3}$ (considering interactions for which the strong coupling scale is $\Lambda_5=(m^4 M_W)^{1/5}$ would lead to a much weaker bound on $g_*$ from the improved positivity bounds alone $g_*<(m/M_W)^{1/20}$ but such interactions do not preserve the positivity bounds beyond the forward limit \cite{deRham:2018qqo}).

When applied to a theory of massive gravity, $M_W=M_{\rm Planck}$ and (even though in principle possible) we would not expect the graviton mass to be parametrically much larger than the Hubble parameter today \cite{deRham:2016nuf}. Further demanding for a Vainshtein mechanism to take place while remaining in the weak coupling \eqref{eq:g*} sets a value for the cutoff to be \cite{deRham:2017xox}
\begin{eqnarray}
\label{eq:cutoff}
\Lambda_{\rm cutoff} \lesssim 10^{-5} \Lambda_3\,.
\end{eqnarray}
This is a typical example of how requirements from a standard UV completion demand new physics to enter at a scale well-below the strong coupling scale $\Lambda_3$ even though from a  pure low-energy approach perturbative unitarity only starts breaking down at scales close to $\Lambda_3$ and there would appear {\it no need} for new physics to enter well below that scale.

Within the current state-of-the art the bound \eqref{eq:cutoff} is the most stringent bound one has been able to impose on the cutoff of  a massive gravitational theory. Interestingly this bound relies on the requirement that the UV completion remains:\\

\hspace{2cm}\begin{minipage}{8cm}
\begin{itemize}
\item Weakly Coupled,
\item Strictly local, Lorentz invariant and unitary.
\end{itemize}
\end{minipage}\\[0.3cm]

In principle any of these assumptions could be violated. For instance in models like massive gravity and bigravity, relaxing the weak coupling requirement may eliminate the need for new physics at a low energy scale. Moreover relaxing the strict locality assumptions also opens up possibilities \cite{Keltner:2015xda} that may deserve further explorations.\\

A yet different set of criteria used to constrain massive spin-2 theories solely based on theoretical considerations is that of asymptotic (sub)luminality as was considered in \cite{Camanho:2016opx,Hinterbichler:2017qyt,Bonifacio:2017nnt,Hinterbichler:2017qcl}, which impose constraints on cubic operators of massive spin-2 fields. Roughly speaking, these criterion demand that the Eisenbud-Wigner scattering time-delay \cite{Wigner:1955zz} is positive, encoding the causality of the scattering set-up. In \cite{Hinterbichler:2017qyt,Bonifacio:2017nnt} it is argued that in order to preserve asymptotic (sub)luminality, the cutoff of the EFT for a spin-2 field is $\Lambda \sim m$ unless the cubic interactions are tuned into a special form that corresponds to a one-parameter family of the ghost-free massive gravity models \cite{deRham:2010kj}. Once this technically natural tuning is made, the cutoff reverts to the standard one as above.

\section{Discussion}

The existence of a cut-off for EFTs of massive spin-2 particles which acts as a quantum gravity {obstruction} to taking the limit $m \rightarrow 0$ is nothing new. In the context of ghost-free models of massive gravity, the strong coupling scale is known to be at most $\Lambda_3 \sim (m^2 \mpl)^{1/3}$ with possible new states arising at scales below this.  Since $\Lambda_3 \rightarrow 0 $ as $m \rightarrow 0$ for fixed $\mpl$ we guarantee an obstruction to taking the massless limit. The crucial question is what is the highest cut-off scale that allows a Lorentz invariant UV completion, i.e. which does not belong to the Swampland. The conjecture of {\cite{Klaewer:2018yxi}} is that this scale is the {parametrically} lower scale {$\Lambda_m  \sim m \mpl/M_W$} and in the case of a nonlinear theory of massive gravity $\Lambda_m \sim m$. While these conjectures may prove true for other reasons, the arguments and examples given in {\cite{Klaewer:2018yxi}} are themselves spurious and do not support the conjecture. The application of the WGC to the helicity-1 mode of the massive spin-2 state does not stand up to scrutiny since there are no {meaningful} charged states which source it, and its contribution can never be {isolated from} the dominant attractive contributions from the helicity-2 and helicity-0 modes.

By contrast, the minimal form of WGC we may actually expect to apply for a massive spin-2 particle is simply that the scale $M_W$ governing the interactions of the massive spin-2 field be
\be
M_W<\mpl\,,
\ee
meaning that the force exchanged by the massive spin-2 particle dominates over that for the massless at distances less than the Compton wavelength. Interestingly this conjecture holds in all multi-gravity theories, such as bigravity and its massive gravity limit. The reason it holds is that the metric with the largest Planck Mass is always the one that is dominated by the massless graviton mode. In addition the kinetic terms for the massless graviton add coherently and positively.

The strongest bounds on the cutoff of interacting spin-2 fields come from the application of both forward and non-forward limit positivity bounds \cite{Cheung:2016yqr,deRham:2017xox,deRham:2018qqo,deRham:2019wjj}  and the improved positive bounds \cite{Bellazzini:2016xrt,deRham:2017imi,deRham:2017xox} which account for loop corrections. It is likely the bounds of this type have not yet been exhausted, and a more stringent form of them may further lower the cutoff of these EFTs. A technical challenge in considering these bounds for massive states coupled to gravity is that the massless graviton $t$-channel exchange dominates in the vicinity of the forward limit. Related works on asymptotic (sub)luminality explore a different regime of the S-matrix  \cite{Camanho:2016opx,Hinterbichler:2017qyt,Bonifacio:2017nnt,Hinterbichler:2017qcl}. A far less well explored possibility is what happens if the UV completion is not weakly coupled.\\

\noindent{\textbf{Acknowledgments:}}
We would like to thank Lasma Alberte, Costas Bachas, Sergio Ferrara, Alex Kehagias, Daniel Klaewer, Dieter Lust, Eran Palti and Angnis Schmidt-May for very useful and interesting discussions. CdR and AJT thank the Perimeter Institute for Theoretical Physics for its hospitality during part of this work and for support from the Simons Emmy Noether program. The work of CdR and AJT is supported by an STFC grant ST/P000762/1. CdR thanks the Royal Society for support at ICL through a Wolfson Research Merit Award. CdR is supported by the European Union's Horizon 2020 Research Council grant 724659 MassiveCosmo ERC-2016-COG and by a Simons Foundation award ID 555326 under the Simons Foundation's Origins of the Universe initiative, `\textit{Cosmology Beyond Einstein's Theory}'.   AJT thanks the Royal Society for support at ICL through a Wolfson Research Merit Award. LH is supported by the Swiss National Science Foundation.
\bibliographystyle{JHEP}
\bibliography{references}

\end{document}